\documentclass[aps,twocolumn,amsmath,amssymb]{revtex4}

\usepackage{graphicx}
\usepackage{dcolumn}
\usepackage{bm}
\usepackage{amsmath}
\usepackage{fancyvrb}

\begin{document}

\title{Microcontroller interrupts for flexible control of time critical tasks in experiments with laser cooled atoms}

\author{Mark Sadgrove} 
\affiliation{Institute for Laser Science, The University of
Electro-communication, 1-5-1 Chofugaoka, Chofu, Japan}

\begin{abstract}
I detail applications of timer interrupts in a popular micro-controller family to time
critical applications in laser-cooling type experiments.
I demonstrate a low overhead 1-bit frequency locking scheme and a multichannel experimental
sequencer using the timer-counter intterrupts to achieve accurate timing along with flexible 
interfaces. 
The general purpose nature of micro-controllers can offer unique functionality compared with commercial solutions
 due to the flexibility of a computer controlled interface without the poor latencies
 associated with computer timing.
\end{abstract}


\maketitle

\section{Introduction}
\subsection{Motivation}
Novel experiments often have control requirements which
fall outside the parameters offered by commercial software and
hardware solutions. Additionally, the proprietary nature of 
commercial hardware can present problems when it comes to accurately
characterising and extending the hardware used for
experiments. For this reason, the continued rapid progress of 
integrated circuit technologies is important, providing more and more
speed and functionality for a given volume  with each passing year.
One intriguing use of the general purpose functionality of modern integrated circuit
technologies is to provide custom control of experiments.

A case in point is microcontrollers. A number of these cheap, popular 
``single chip computers'' now have roughly the same speeds and 
memory sizes as personal computers of two decades ago. However,
 unlike a standard personal computer where useage is mediated by an operating
system (OS),
almost all the computational resources of a microcontroller are available to be
used at the discretion
of the user.
The power and flexibility of modern microcontrollers has already been put to use
in a number 
of experimental settings of which I give some recent examples in
Refs.~\cite{miconexp}.
In the dynamic field of laser cooled atom based research, a recent work has 
amply showcased the flexibility of microcontrollers along with open
source software to create a unique control system for a cold atom 
experiment~\cite{Steck}.

Here, my aims are somewhat parallel  to those of
Ref.~\cite{Steck} but 
I focus on a more specific aspect of micro-controllers - that of \emph{interrupts}. The 
principle message of this paper is that the interrupt features of modern microprocessors (described below)
offer a convenient alternative to commercial hardware and software to achieve control of experiments where strict timing is required 
along with flexible interfacing for the user.

An additional advantage is that microcontrollers offer an entry-level way to introduce students to 
programming and electronics for experimental control. Although I focus on a laser cooling and atom interferometry related application here,
the control of timing, frequency and phase of fields covered by these devices are vital aspects across a broad range of physics experiments.

Interrupts are a standard feature of modern microcontrollers.
Essentially, an interrupt is a processor function whereby a code branch which is executed as soon
as a certain variable changes state. In principle, the variable
may be a hardware register or a software variable proper, although
in the present paper, I consider only interrupts generated
by hardware registers of a microcontroller. Using this method, it is possible to
create flexible devices with $\sim 10\mu$s worst case latencies -- 
good enough to control modern atom interferometer experiments, for example.


The purpose of an interrupt is to guarantee the execution of a code
branch regardless of where in a software program a microprocessor may
currently be. This functionality immediately lends itself to the 
creation of timer triggered  and external triggered events, both 
of which are essential in laboratory equipment, but are rarely 
available in more general purpose devices. It is fair to say that
such microcontroller interrupts are essentially a method to provide hard real
time sequencing. However, in contrast with other methods of provding
high-performance timing, such as polling of a given input channel, 
interrupt based timing allows for a richer software interface to be created since the processor may execute
additional code (e.g.a user interface) until the interrupt is received. For example,
in the devices I present below, sophisticated computer interface routines 
with the microcontroller were possible while retaining timing accuracy.
This fact allows for the construction of genuinely flexible real time instruments necessary
for the control of many physics experiments.

Here, I present two applications 
of microcontroller interrupts   - frequency locking and multichannel sequencing. 
Both are realized using the 
same microcontroller family~\cite{ATM168DataSheet,ATM1280DataSheet} running at a 16MHz clockspeed.
A popular universal serial bus (USB) programmer called Arduino~\cite{arduinoref} was used for programming 
the microcontrollers (using the c language) and
to allow a serial computer interface for control. Although I chose to use the
same microcontroller family for each application, interrupts are a generic
feature of microcontrollers (see, for example~\cite{PIC} and~\cite{ARM}), and the results
given here are in no way limited to specific micro-controller hardware.

\subsection{An inexpensive, popular microcontroller programmer: Arduino}
In pursuing the aims discussed above, I chose to use a popular microcontroller 
programming board called Arduino~\cite{arduinoref}. The Arduino project offers
a universal serial bus (USB) interface coupled with a programming environment
which allows
simplified building and uploading of C code to popular Atmega microcontrollers
made by the company Atmel.
In particular, here I used the Arduino Duemilanove and Arduino Mega boards
which correspond to the Atmega
168 and 1280 microcontrollers respectively.

Although the Arduino project is intended for hobbyists or people who
have little
confidence or experience with microprocessors, in fact it merely offers a
simplified interface 
to programming the Atmega chips with standard AVR C\footnote{AVR C is the name
of the C language standard created by Atmel
to allow C programming of their micro-controllers.}, and therefore Arduino in
principle 
allows full use of all the microcontroller's functions. Indeed, in the applications
considered here,
I directly set microcontroller registers and use timer interrupt functions
which lie
outside the documented purview of the Arduino project, but are nonetheless
easily implemented
in the Arduino programming environment.

Because of the full access to functionality inspite of the simplified
presentation, I believe the
Arduino project is also suited to physics laboratories, particularly in the case where
students may wish
to build micro-processor projects but have slim electronics or programming
experience.
\section{The experiment}

The microcontroller applications I consider here were designed for a 
``quantum control" experiment involving cold atoms in which atoms were subject to
a phase shifted potential (also controlled by a microcontroller~\cite{DDS}) 
in order to control such effects as quantum localization~\cite{Super}. Because the experiment
involved a number of pulses, each of which was separately phase controlled,
it is in many ways similar to an atom interferometer experiment, and I will simply
refer to it as an ``atom interferometer experiment" for simplicity.

I now briefly describe the experimental setup. 
A standard six-beam magneto-optical trap
(MOT)~\cite{monroe} of Rb87 was
built using two frequency stabilised external cavity laser diodes (ECLD). For
laser cooling, the frequency 
stabilization was performed using a microcontroller to create a servo loop lock
to a saturated absorption signal.
This scheme will be described in Section~\ref{Sec:FreqLock} below. A repumping
laser was frequency stabilized by a servo system at 6.8 GHz from
the cooling laser. This was achieved by prescaling the beat signal between the repump and cooling
laser and using a frequency 
phase/frequency discriminator to give an error signal based on the
difference between the downshifted beat signal
and a local oscillator.

After amplifying the cooling laser using a tapered amplifier, repump and cooling
beams were overlapped, split in 3 and 
sent through orthogonal ports of a stainless steel vacuum cell with appropriate
circular polarizations. Retroreflecting the
three beams (again using waveplates to ensure the correct polarization) created
an optical molasses, and application of a 
quadrople field  gradient using anti-Helmholz coils produced a MOT.

We also  used a second microcontroller to create a 32 channel experimental sequencing
device to control the various optical
and magnetic fields in the experiment and make possible atom-interferometer type
experiments. The specifics
of this sequencer are described in Section~\ref{sec:Seq} below. Using the
sequencer to create a polarization gradient cooling (PGC) event sequence, the atoms were cooled
 further to
below 20$\mu K$ and subsequent application of standing wave pulses
to the atoms and data acquisition was also controlled.

Finally, in order to perform interferometric type experiments with the atoms, a
standing wave was created from two counterpropagating
beams which could be independently frequency tuned using acousto-optic
modulators (AOMs). The AOMs were driven
by the amplified signals from two phase synchronised direct digital synthesis
(DDS) devices.
We also used a micro-controller to allow the phase of one of the beams to be changed
between preselected values by the 
application of an external trigger signal, the method for which is described elsewhere~\cite{DDS}. 

\section{Frequency locking and multi-channel sequencing}
The two  applications of micro-controllers considered here are generally considered to 
be solved problems in the atom-cooling community. The frequency stabilization of a laser to a
saturated absorption signal peak is usually accomplished using frequency modulation and 
a lock-in amplifier, whilst multi-channel sequencing of experimental equipment is typically
accomplished with commercial hardware and software.
Nonetheless, I show below that micro-controller interrupts coupled with basic auxillary circuits
offer a  relatively simple and more flexible approach to both of these problems. \\

\subsection{1-bit frequency stabilization device}
\label{Sec:FreqLock}
\begin{figure}
\centering
\includegraphics[width=\linewidth]{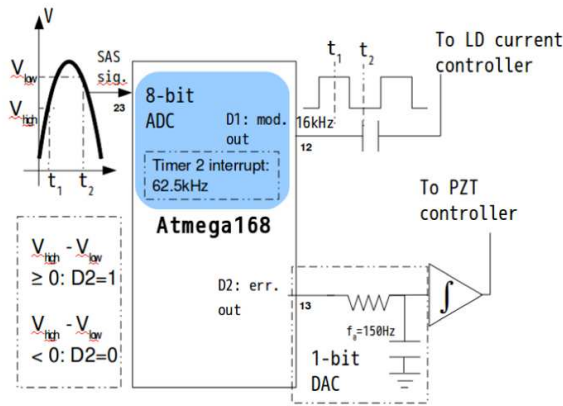}
\caption{\label{fig:Lock} Diagram of microcontroller frequency locking setup. The 
laser diode frequency is modulated by a digital signal, and the resultant change
in the SAS signal is sampled synchronously giving $V_{\rm high}$ and $V_{\rm low}$.
A 1-bit error signal is generated asynchronously by comparing the signals and this error 
signal is low-pass filtered and sent to an integrator before being sent to the piezo controller.
 }
\end{figure}
In this section I focus on a novel peak lock system with the following notable characteristics:
\begin{itemize}
\item 1-bit laser frequency modulation.
\item 1-bit digital-to-analog (DAC) error signal.
\end{itemize}

I emphasize that, while digital frequecy stabilization is not a new idea (see, for example,~\cite{digilock}),
the simplicity of the scheme considered here (in its use of a single microcontroller and purely digital outputs) is
 a strength. For example, the use of 1-bit signals to provide modulation and error signals
means that the method is suitable for implementation on microcontrollers with only a 
few outputs and no onboard DAC. Thus, I refer to the system  as \emph{low overhead} in terms
of the resources it requires to function.

\subsubsection{Some background on laser frequency stabilization}
To cool atoms using lasers, it is necessary to accurately and stably lock the
frequency of a laser beam at 
a set detuning from an atomic cycling transition. This feat is achivable using
saturated absorption 
spectroscopy (SAS) the peaks of which spectrum, as detected by a photodiode,
provide an electronic
 reference signal which may be used to create a servo loop in which variation of
laser diode current and/or
the position of an external feedback grating (by piezo-electric means) keeps the
frequency of a laser stable to within MHz.

The details of the servo-locking scheme are important as they determine the
``strength" and reliability of the lock.
For example, the simplest way to lock to an SAS peak is to use the midpoint of
one side of the peak as a zero reference.
Frequency changes of the laser lead to movement of the peak and thus induce
differences in the photo-diode voltage from the 
reference voltage. This difference can be integrated over time and used directly
as an error signal to correct the
frequency of the laser.

However, this simple scheme ignores some standard laboratory realities - namely
that changes in the photodiode voltage
are not always due to frequency changes in the laser and may be due to gradual
alignment shifts which decrease the coupling
of the SAS probe light to the photodiode. For this reason, sidelocking
techniques are unstable over time.
This has lead to peak locking techniques gaining favour in the laser cooling
community. 

Peak locking requires more sophisticated electronics than the side-locking
scheme. To begin with, frequency modulation of the 
laser diode output must be arranged. This is typically acomplished either by
modulating the laser current directly or by 
modulating the output beam with an electro or acousto-optic modulator. The SAS
probe absorption signal is monitored in phase
with this modulation. Filtering the demodulated
absorption signal can provide a signal which is effectively the derivative of the 
SAS peaks, thus providing a stable zero crossing at the peaks which can be
locked to. This derivative signal is stable against low
frequency drifts in the SAS signal. Although by no means difficult technically,
the modulation and demodulation steps 
described above typically require more sophisticated electronics knowledge to
contstruct and are performed by analog 
modules for modulation and demodulation whose operation is opaque.
 It is this fact that led me to design a very simple peak locking
system in which both modulation and locking
are performed \emph{all-digitally} by an Atmega 168 micro-controller. 


\subsubsection{1-bit peak locking method}
We used an 8-bit counter/timer interrupt~\cite{ATM168DataSheet} routine
with minimal auxillary electronics to lock the frequency of a laser 
as follows (see Fig.~\ref{fig:Lock} and Table~\ref{tab:lockcode}): 
The micro controller executes a timer interrupt routine every 65.5kHz.
The routine toggles between laser current modulation and sampling of the absorption signal.
A \emph{digital} frequency error signal is derived from the \emph{sign} of the difference 
between the signals at high and low current modulation. This error signal is filtered and 
fed back to the PZT driver input. 
The addition of a 150Hz low pass filter at the error output creates an effective 
1-bit DAC. (This is the same from of DAC used in many CD players, although here, the output here is offset so that it swings between -2.5V and 2.5V).
Given that the time constant of the filter is $\sim1$ms and the error signal can be updated as fast as 32kHz,
the effective granularity of the filtered error signal is 32 distinct voltage levels (5-bit).
 The micro-controller pseudo-code for the method described here is
given in Table~\ref{tab:lockcode}.

This peak-lock method is a simplified version of standard current modulation
techniques. Instead of demodulating the
SAS signal to produce a derivative of the spectrum, the difference in voltages
measured for high and low digital
modulation are used to indicate which side of the peak the current laser
wavelength is. This crude approximation to
the derivative is good enough for an error signal because it is fed through a low pass filter followed by an
integrator which smooths the digital signal and ensures that the error signal fed to the
piezo is proportional to how far from the
peak the laser wavelength is. 

Because 65.5kHz interrupt function is split between modulation and 
servo adjustment, the Nyquist limited maximum update frequency of the servo (i.e. the feedback
bandwidth) is 16kHz (2s.f.).
At present only feedback to the piezo is used which is adequate for current
experimental purposes. The lock is stable against any drifts in the SAS signal,
and typically lasts for an
entire day of experiments without requiring relocking. However, without current
feedback, sharp shocks, such as those created by dropping
an object on the optical table where the laser is mounted, are enough to destroy the lock. Applying current
feedback should remedy this situation,
although the upper frequency limit of $\lesssim16$kHz available for
current feedback signals is somewhat
lower than that available with popular analog amplitude modulation techniques
and may not afford the same protection against 
sharp shocks. 
Although the locking device was not connected to a computer in the present application,
in principle, the locking circuit could be used with a USB programmer~\cite{arduinoref}, allowing a
computer to monitor and control the locking of the cooling laser. Automatic relocking and active
feedback gain adjustment are certainly realistic extensions of the current method. 

As an amusing aside, I note that the code used here is adapted with permission from a
project from the Centre for Media Arts in
Cologne, which used the interrupts to provide sampling of an audio signal in
order to produce a guitar effect~\cite{reverblink}.
In the repurposed code, which may be found at the link given
in~\cite{codelinklock}, I have also added a large number of detailed comments
in the interest of providing an accessible introduction to the use of microcontroller
interrupts for instrumentation purposes.

\begin{table}
\begin{center}
\end{center}
\begin{Verbatim}[frame=single,framerule=0.2pt,framesep=5pt] 
loop {
  if sample == 1 do {
    e = Vhigh - Vlow
    if e < 0 write 0 to pin 9
    else write 1 to pin 9
    sample = 0 }
}

timer interrupt handler {
  toggle f
  if f is 0 do {
    write 0 to pin 6
    wait 10 micro seconds
    Vlow = voltage from ADC }
  else do {
    write 1 to pin 6
    wait 10 micro seconds
    Vhigh = voltage from ADC
    sample = 1 }
}
\end{Verbatim}
\caption{\label{tab:lockcode} Pseudo-code for the frequcency lock
micro-controller application.}
\end{table}

\subsection{A flexible experimental event sequencer}
\label{sec:Seq}
\begin{figure}
\centering
\includegraphics[width=\linewidth]{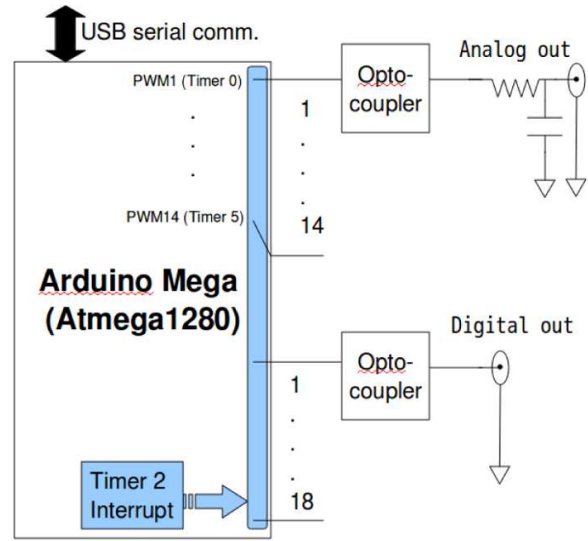}
\caption{\label{fig:ScaleRes} Diagram of the experimental sequencer. 14 analog
and 18 digital channels are available.
Both analog and digital output stages begin with an opto-coupler to isolate the
microprocessor side from the instrumentation
side. The analog output stage uses the PWM enabled pins of the Atmega1280 and
thus requires lowpass filtering before
use. }
\end{figure}
I now consider an important application for many different types of laboratories: real
time experimental sequencing. The sequencing device described below has the following features:
\begin{itemize}
\item 14 analog and 18 digital output channels.
\item Sequencing of up to 256 discrete events involving arbitrary numbers of channels.
\item $\sim10\mu$s latency of event timing over long ($\sim24$hr) time scales.
\item Simple opto-isolation of all digital and analog channels.
\item Computer controllable by standard serial communication (USB).
\end{itemize}
The last of the above characteristics may seem trivial, but it allows some interesting control possibilities.
For example,  I was able to control the experiment completely from a cellphone
by sending commands over a secure shell connection. Far from being a gimmick,
such a feature provides a flexible method for multiple users of an experiment to 
have remote control over it. This is useful, for example, in 
alignment or testing phases of an experiment, where fields may need to be turned off and on 
when the user is standing at a location separated from the main control computer.

\subsubsection{Using microcontroller interrupts for real time control}

In practice, ``real time" means that the worst-case latency in timing (that is
the difference between the time
we set an event to occur at and the time at which it actually does occur) is
sufficiently small that it does not 
affect the reliability of the experiment. In the case considered here, atom interferometer type
experiments typically require 
a time-of-flight (TOF) technique to measure the final atomic position or
momentum distribution. Accurate timing of 
the TOF sequence requires synchronized switching of laser and magnetic
fields along with triggering of a 
measurement device (typically a charge coupled device (CCD) camera). A typical
TOF measurement is conducted 
over 10ms and thus events during the TOF sequence must have latency and timing
jitter which are small compared
with this time scale.

In the experiment considered here,  less than $1\%$ latency was achieved on the 10ms time scale for
32 independent, programmable channels using a microcontroller~\cite{ATM1280DataSheet} controlled by a
USB programmer. Although I controlled the sequencer using a computer running the 
Linux operating system, the USB interface is generic, and the sequencer can be
programmed from any computer
which supports USB. Control sequences may be uploaded by sending standard serial
control signals to the microcontroller which is programmed with custom sequencer software~\cite{seqcodelink}.

The sequencer operates in three separate modes, all of which are controlled from the computer
by standard USB serial communications. The modes are: (i) \textbf{normal mode} where
analog channel values may be set one at a time by the user sending
the appropriate serial commands,
(ii) \textbf{event entry mode} where analog or digital 
channel/value pairs are added to an ``event" which occurs at a specific time in
the experimental schedule and 
(iii) \textbf{sequencer mode} in which the events in the experiment schedule are
exectuted in sequence controlled
by a timer overflow interrupt~\cite{ATM1280DataSheet}.

This method of timing experimental events in the sequencer mode stands apart from
many commercial methods in which events are essentially encoded into random access memory
and then read out at a constant clock rate. The sequencer mode used here is similar to that found in~\cite{Steck},
and due to the long integer format used for timing allows timing of events down to $\sim10\mu$s
acuracy over the course of a whole day. This means that, aside from the application I consider 
here, such a method should be useful in astronomy experiments or for control of data-logging 
which requires good timing accuracy over many hours.

Pseudo code for the major functions of
the sequencer is given in 
Table~\ref{tab:seqcode}. The full code, which includes extensive comments and
references to the relevant hardware,
may be downloaded from the link
given in Ref.~\cite{seqcodelink}
\begin{table}
\begin{center}
\end{center}
\begin{Verbatim}[frame=single,framerule=0.2pt,framesep=5pt] 
loop {
  opcode = serial input 
  n = serial input
  param1 = serial input 
  :
  :
  param4 = serial input
  if (opcode == a or d) and 
	  mode == normal  do {
    set channel n to value given by param1 }
  if (opcode == a or d) and 
	  mode == event_entry_mode do {
    Add {n, param1} to the current event }
  if opcode == e do {
    mode = event_entry_mode
    make a new event in the event shedule
    set the time of the event to n }
  if opcode == f do {
    mode == normal }
  if opcode == x do {
    mode = sequencer
    enable timer interrupt }
  
  if mode == sequencer do {
      if t == next_event_time do {
	loop through CV pairs in event }
      if all events done do { 
	disable timer interrupt
	t = 0
	mode = normal } }
}

timer interrupt handler {
  increment t
}
\end{Verbatim}
\caption{\label{tab:seqcode} Pseudo-code for principle functions of the
experimental sequencer.}
\end{table}

\subsubsection{Digital and analog outputs}
32 channels are available for independent control. These channels are divided
into 18 digital and 14 analog
channels. 
\emph{Pulse width modulation} (PWM) was used on 14 digital outputs which, after filtering,
provided an effective 
8 bit output between 0 and 5 volts. This has the advanatge that the
same opto-coupled output stage can be used as for the digital outputs
with extra filtering for the analog channels after the
opto-couplers.  I used Sharp PC900V optocouplers
for all of the outputs which inverted the output of the Atmega micro-controller.
The optocouplers give rise-times
on the digital outputs of $\sim 200$ns, which, although much longer than
standard TTL digital rise times, was more than
short enough for my purposes.

PWM works by tying the value of an 8 bit timer register to a digital
output pin by using the \emph{output compare register} (OCR)
method~\cite{ATM1280DataSheet}. The timer register  which is incremented each clock cycle, 
is compared with an 8-bit OCR register whose value can be set by the user.
 When the timer value matches the OCR value, the digital output is flipped. By
lowpass filtering this modulated digital
output, a smooth quasi-DC output with 256 levels is achieved. PWM does
have some 
disadvantages compared with real DAC. The underlying PWM frequency is the
clock-speed/256  = 62.5kHz and the filtering 
circuits needed to smooth the output must operate at a value of at most half this
frequency. This fact, coupled with software issues specific to the microcontroller I used~\cite{ATM1280DataSheet}
 means that the analog channel rise time
is significantly longer than the digital output rise time (See table~\ref{tab:timing}).

I note that there is some residual ripple on the analog output (see Fig.~\ref{fig:SeqTrace}). This could be removed 
by using an output stage filter with a lower frequency cutoff at the expense of
increasing the rise time of the analog channels. However, I found that the analog outputs
were sufficiently noise-free to perform the operations required in the laboratory. For example,
the analog outputs were used to control the power and detuning of the cooling laser in the experiment
in order to effect sub-Doppler cooling, and were routinely able to reach sub-20$\mu$K temperatures, 
a result which is sufficient for experiments and comparable with results achieved using commercial
digital-to-analog conversion boards. 

\subsubsection{Timing performance}
\begin{figure*}
\centering
\includegraphics[width=0.9\linewidth]{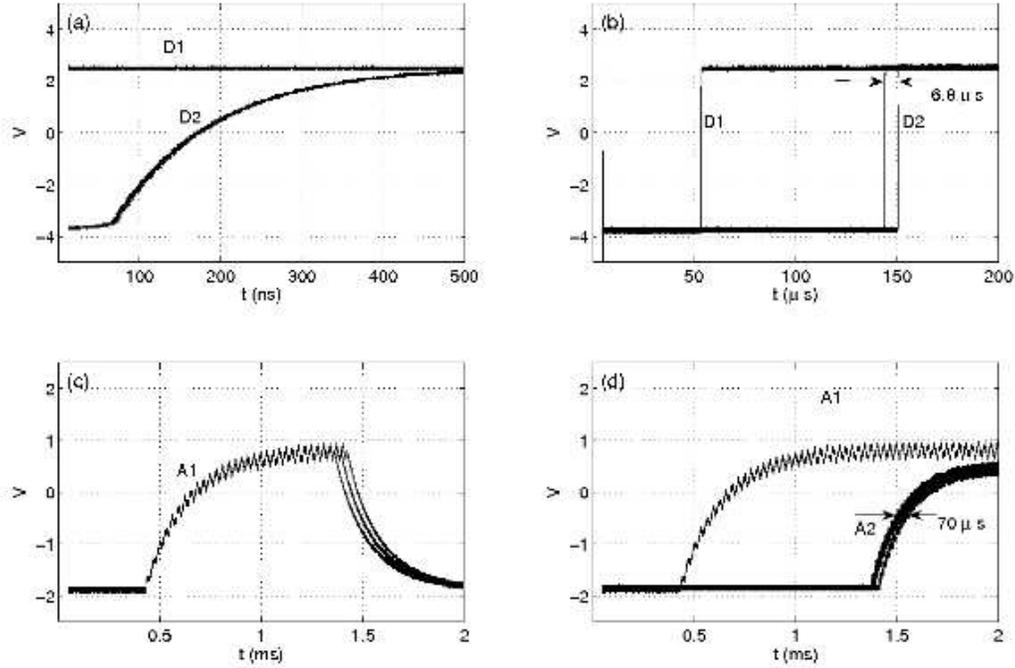}
\caption{\label{fig:SeqTrace} Measurements of the timing accuracy of digital and
analog signals of the sequencer.
(a) Shows 20 traces for the digital output channel 2 (D2) as triggered by channel 1 (D1).
In (b) the systematic
timing error is shown (indicated by the arrows) of the digital channel. The area
indicated by the arrows shows 20 traces of
channel 2 as triggered by a channel 1 rising edge. The channel 2 signal rising
edges gather around one of two distinct
times. In  (c), 20 traces of a pulse from analog channel 1 (A1). Here also, a
systematic timing error leads to 
falling edges of the pulse gathering around 3 distinct times. Lastly, in (d) 20
traces of a channel 2 (A2) rising edge, 
triggered by the channel 1 rising edge, are shown. The maximum horizontal
thickness of the line, indicated by the two 
arrows, gives the timing uncertainty.}
\end{figure*}

Fig.~\ref{fig:SeqTrace} shows traces from four oscilloscope measurements used to
gauge the timing accuracy of the sequencer.
Fig.~\ref{fig:SeqTrace}(a) and (b) are measurements of the outputs of two
digital channels. In (a), the same sequence 
was run 20 times in persist mode (so that all 20 instances are overlaid on the
same trace). The signal consisted of 
 digital channel 1 of the sequencer going high as the first event and digital
channel 2 going high as the second event
1000 interrupt cyles later. The ocscilloscope was triggered off the channel 1
signal. I used the data shown in (a) to
determine the jitter in timing by measuring the width of the 20 overlaid channel
2 signal traces. The thickness
at half maximum of the traces was 7.8ns (about 13$\%$ of the 16MHz clock
period) which gives a
 conservative estimate of the timing jitter on a digital channel. For analog
channels in similar experiments (shown in
Fig.~\ref{fig:SeqTrace}(c) ) the thickness of the acumulated traces was
30$\mu$s.
 By contrast, the time between nominally simultaneous events created by 
the sequencer is 6.2$\mu$s for digital channels or 17$\mu$s for analog
channels and the respective rise times 
of digital and analog channels are 240ns and 360$\mu$s. 

In principle, use of interrupts can provide exactly repeatable hard real time
sequencing; that is, ignoring the 
time between nominally simultaneous events (i.e. the systematic offset to the
timing using the method of executing events one by one from a list) 
only the timing jitter should affect the real time latency. However, because
interrupts work by \emph{literally interrupting}
the main code running on the micro-controller and branching to the interrupt
handler, the details of the code do have an effect
on the timing. This effect was seen, as shown in Fig.s~\ref{fig:SeqTrace}(b)
and (d) for digital and analog outputs
respectively, in a slight \emph{systematic} difference between individual
realizations of a sequence. For the 
digital channels, as shown in (b), the timing of channel 2 going high fell in to
two distinct groups separated by
6.8$\mu$s. This value is the worst case latency of the digital channels: that is
if the user specifies the timing of one
event relative to another, the difference between nominal and actual event times
can differ by up to 6.8$\mu$s on the digital
channels.

Not surprisingly, the latency is worse for the analog channels as shown in (d).
Here the 20 traces of the channel 2 output,
as triggered by the rising channel 1 ouput, fell into three distinct groups,
although with the larger timing jitter, only a thick 
line is visible in (d). The thickness of this line at half maximum was 70$\mu$s
which is the worst case latency for the
analog channels.
\begin{table}
\begin{center}
\begin{tabular}{|l|l|l|}
\hline
&\textbf{Digital} & \textbf{Analog} \\
\hline
\textbf{Rise time}& 240ns &  370$\mu$s\\
\textbf{Timing jitter} & 7.8 ns & 30$\mu$s\\
\textbf{Worst case latency} & 6.8$\mu$s & 70$\mu$s  \\
\textbf{Time between nominally} &&\\
\textbf{simultaneous events} & 6.2$\mu$s & 17$\mu$s \\
\hline
\end{tabular}
\end{center}
\caption{\label{tab:timing} Timing parameters for analog and digital channels of
the sequencer. Values are given to 2 s.f. accuracy.}
\end{table}
To put the channel latencies in perspective, it is necessary to compare them with the
usual TOF measurement period, which is 10ms.
Even the analog channels still have a latency less than 1$\%$ of this value, so
we can see that the sequencer
will provide reliable timing for typical atom interferometer type experiments.
Important timing uncertainties for 
digital and analog channels are summarised in Table~\ref{tab:timing}.

\subsection{A brief note on the computer software interface}
Although development of microcontroller based hardware is the principle
subject of this paper, the software system used to interface the 
sequencer and control the experiment is worthy of a brief discussion.
I created a custom software graphical user interface (GUI) to control the
sequencer along with the rest of the experiment.
The principle software was written in the C programming language, and run on an
IBM Thinkpad using Ubuntu Linux.

\begin{figure}
\centering
\includegraphics[width=0.9\linewidth]{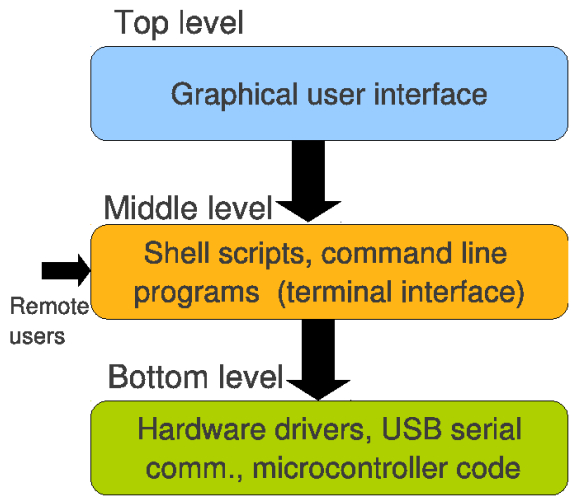}
\caption{\label{fig:Soft} Schematic diagram of the three main software layers used to control the experiment.}
\end{figure}

Typically, the control system (top level) for an experiment is a graphical user interface (GUI) which interacts
directly with hardware drivers (bottom level) to control the exeriment. 
Here, however, I introduced a \emph{middle level} of software which consisted of 
scripts and command line programs which could all be run from a terminal.
This allows much more flexibility when it comes to running the experiment remotely or 
from multiple users' computers. The GUI sits on top of the middle layer,
providing merely a convenient interface rather than any complex functionality.
The code archive for the GUI may be found at~\cite{GUIcode}, while the various
scripts and command line programs may be downloaded at the link given in~\cite{scriptscode}.
The structure of the software is illustrated in Fig.~\ref{fig:Soft}

Scripts to program the timing controller were written in the Python 
language, and the BASH shell scripting
language. All peripheral instruments were controlled by USB, including a CCD
camera (Apogee ALTA U260) which was used 
to take data, and an Agilent pulse generator (Agilent 33220A). Experiments
involving arbitrary numbers of events could be designed and then executed automatically by the
software without need for 
human intervention. 

The experimental automation system actually used a separate C program which is
executed  in the background and then controlled using two ``first in first out" (FIFO)
files. As part of the middle level of the software code, this allows the gui to be modified separately from the experimental control code and also
allows any software that can write and read FIFOs to control the system.

One interesting point regarding the control of the sequencer by standard
USB serial commands from the middle level of software is that it makes control
from a terminal very easy. In particular, in an age when many mobile devices are
connected to the internet,
but mobile bandwidth is not sufficient to usefully pipe graphical displays back
to devices, this simple control
method seems to us to be the optimal way to allow remote control of an
experiment. As an example,
it was possible to control all aspects of the experiment from a cell-phone running
the Android operating
system using the ConnectBot software to open a secure shell to the control
computer, and issuing serial commands
to control the sequencer from that shell. Far from being a gimmick, we
believe that this method provides
the most flexible way to provide a flexible remote control to the experiment to
multiple users. 

\section{Summary}

The microcontroller applications considered here
 use only ``off-the-shelf" components 
but they achieve levels of accuracy in timing,
frequency and phase control
which are suitable for running precise interferometer sequences with cold atoms.

The use of micro-controllers rather than commercial hardware solutions (assuming
such solutions even exist) 
enables customised tools to be created for specific experimental situations and
also provides more opportunities 
for learning electronics and programming skills for students compared with
commercial ``black boxes". Indeed,
although the applications here were developed  
for performing and novel experiments,  one of the principle strengths
of using micro-controllers 
to run experiments, particularly when coupled with the easy to use Arduino
programmers, is the  opportunity they afford for education.

The applications considered here would benefit greatly if the specifications of the
micro-controllers were improved. The most useful
improvement would be an increase in the clock speed. This would make current
feedback more viable in the locking application and
improve the worst-case latency of the experimental sequencer.
Additionally, the availability of hardware supported DAC would improve the
frequency lock and sequencing applications.

I also note that at the moment the experiment proceeds in a linear way as
designed on a computer, and does not
make use of the micro-controller interrupts' great flexibility in dealing with
events generated during an experiment.
There is certainly no barrier to modifying the sequencer application to be
triggered off external interrupts.

\subsubsection*{Acknowledgements}
The microcontroller-based work presented here was developed independently but with the support of Nakagwa-laboratory at
the Institute for Laser Science.
Furthermore, with respect to the entire experiment for which the applications presented here
were created, and also for suggestions regarding the interpretation of the
frequency locing circuit's function, I am very grateful for the advice and support
I received from Ken'ichi Nakagawa.

Additionally, the Arduino project community internet forum and related websites were an invaluable reference
during the undertaking of this project.

\end{document}